\documentclass[aps,prb,floats,twocolumn]{revtex4}
\usepackage{epsfig}
\renewcommand{\vec}{\mathbf}
\begin{document}         
\title{Theory of field induced spin reorientation transition in thin Heisenberg films} 
\author{S.~Schwieger, J. Kienert, and W.~Nolting}
\affiliation{Lehrstuhl Festk{\"o}rpertheorie, Institut f{\"u}r Physik, Humboldt-Universit{\"a}t zu Berlin, 
  Newtonstr. 15, 12489 Berlin}
\begin{abstract}
We consider the spin reorientation transition in a ferromagnetic
Heisenberg monolayer with a second order single ion anisotropy as a
function of temperature and external field. Up to now analytical methods
give satisfying results only for the special case that the external
field is aligned parallel to the easy axis of the crystal.
We propose a theory based on a generalization of the Callen decoupling, which can be used for
arbritrary direction of the external field. Excellent
agreement between our results and Quantum Monte Carlo data is found for
the field induced reorientation at finite temperatures.  
Additionally, we discuss the temperature dependence of the transition in
detail.
\end{abstract}
\maketitle
\section{introduction}
Since the discovery of the Giant Magneto Resistance (GMR) effect 1989
\cite{BGS89} there has been enormous interest and research
activity in the field of thin magnetic films. The magnetic anisotropy,
merely a small perturbation in a bulk ferromagnet, gets strikingly
important in thin film systems. Here the anisotropy is not only a
necessary precondition of spontaneous ferromagnetism \cite{MeW66}, but
it 
determines many system properties as, e.g., the dependence of the
magnetization vector or of the spin wave excitation spectrum on an
 applied magnetic field. Additionally, the anisotropy energy is of the same
order of magnitude as the Interlayer Exchange Coupling (IEC)
\cite{Bru95}, which is intimately connected with the GMR effect. Thus an
investigation of these effects has to take the magnetic anisotropy
carefully into account.\\ There is an important
phenomenon in thin ferromagnetic films which is closely connected to
the magnetic anisotropy: the magnetic reorientation transition. This
term denotes a rotation of the magnetization from the film normal into
the plane or vice versa as a function of temperature, film thickness, or
external magnetic field. The transition can be understood as a result of
competing forces that favor different directions of the magnetization as
e.g. spin orbit coupling, dipolar interaction, and an external magnetic
field\cite{JeB98}. 
It can be described using a Heisenberg model in film geometry, with
 the usual Heisenberg exchange interaction, an external
field, and one or more anisotropy terms.\\
For the simplest of these models,
\begin{equation}
 H_1=-\sum_{ ij }J_{ij}\vec{S}_i\vec{S}_j
-\sum_i \vec{B}_{0}\vec{S}_{i},\nonumber
\label{H1}
\end{equation}
 consisting only of the exchange term and an
external field, there are very accurate approximation schemes available.
It was shown e.g. in Ref. 5 by comparison with QMC calculations, that the
RPA decoupling\cite{BoT59} yields even quantitative results for the
magnetization as a function of temperature.\\
Turning to anisotropy contributions, the spin-orbit coupling induced anisotropy
is usually modelled by a single ion anisotropy 
\begin{equation}
H_2=-K_2\sum_i{S}_{iz}{S}_{iz},
\label{anis}
\end{equation}
which is of second order for systems with tetragonal symmetry. For film
systems the $z$-axis is perpendicular to the film plane.
If $K_2$ is positive the easy axis of the magnetization is the $z$-axis,
for negative $K_2$ this is a hard direction.
The RPA fails badly if applied to a local term as described in
Eq. (\ref{anis}). Thus RPA can not be used to solve the whole model 
\begin{equation}
H=H_1+H_2.
\label{hamiltonian}
\end{equation}
\\
In Ref. 8 an approximation for this model is proposed, which
is based on a combination of the RPA approximation for the nonlocal
terms (\ref{H1})
and an Anderson-Callen (A.C.) decoupling\cite{AnC64} for the local
anisotropy contribution (\ref{anis}). This theory gives good
results\cite{HFK02} for the magnetization if the anisotropy constant $K_2$ is much smaller than the
exchange coupling $J$ ($K_2\le 0.01J$) and if the external field is
applied parallel to the $z$-axis while $K_2$ has to be positive. The
first condition is not a serious restriction, since in reality the
anisotropy constants are indeed much smaller than the Heisenberg exchange
interaction. Furthermore this restriction can be relieved by an
alternative theory\cite{FKS02}. The important restriction is given by
the second condition. The described limit is a very special one, where
both, the anisotropy as well as the external field favor a alignment of
the magnetization parallel to the $z$-axis. Thus there are no
"competing forces", and no magnetic reorientation
transition occurs in this limit, which we want to refer to as "parallel
limit" in the following. However, if the external field is not applied
parallel to the $z$-axis and the magnetization is consequently rotated
out of $z$, the approximation described in Ref. 8 looses
its accuracy and becomes unacceptable for a quantitative
description of the reorientation transition. This was shown in
Ref. 6 by comparison with QMC calculations. Actually, to
our knowledge there is no reliable model theory available, which can treat the model
(\ref{hamiltonian}) for arbritrary directions of the external field or
a negative anisotropy constant $K_2$.\\
However, such a model theory is highly desirable. It can be used to
investigate quantitatively the magnetic reorientation transition in all systems
dominated by  second order lattice anisotropies. Numerical methods, as QMC
calculations, are only applicable for the monolayer, but a finite number
of layers is crucial to study the interplay between surface and bulk
anisotropies (see e.g. Ref. 4).\\
A priori, it is not clear, that such a model theory exists. The full
model (\ref{hamiltonian}) is much more complex then the "parallel limit" case.
The reason is that in that special case the total spin is a conserved quantity
i.e. the total magnetization $\sum_i S_i^z$ 
commutes with the Hamiltonian. This property simplifies the calculations
considerably but it is not present in the general model (\ref{hamiltonian}).\\
In this paper we want to show that nevertheless a well funded model theory can be formulated
and that it is as accurate in the general case as in the "parallel
limit". Comparing our results with the QMC data of reference
\cite{HFK02} we will find excellent quantitative agreement for the
magnetization and its components, which allows
for a high quality description of the magnetic reorientation transition.
The theory proposed in Ref. 8 is recovered for the special
case of the "parallel limit".\\ In this paper we want
to introduce the new theory and evaluate it by comparison with
available QMC results. Since the latter are results for the monolayer,
we will exclusively treat the case of a monolayer during this paper. A
generalization to a multilayer system is straightforward.\\

\section{Theory}
We will assume tetragonal symmetry in the following. The $xy$-plane is the film plane and thus
the 
 $x-$ and the $y-$ direction are
equivalent. That is why we can confine the external field and the
magnetization to the $zx$-plane
without loss of generality. We will assume nearest neighbor coupling
($J_{ij}=J$ for nearest neighbors and $J=0$ elsewhere) and for the
explicit calculations a quadratic lattice. Let us first outline the main
points of our theory. The aim is to calculate the angle and the
norm of
the total magnetization of the
model (\ref{hamiltonian}).
\begin{enumerate}
\item In general, the magnetization is not aligned parallel to the
  $z$-axis. We therefore apply a coordinate transformation that rotates our
  system to align its $z$-axis parallel to the
  magnetization. The calculations are much easier and also more
  convenient in the new system referred to as $\Sigma^\prime$.
\item After this we write down the equation of motion of the single
  magnon Green function. To solve this equation it is necessary to
\item decouple higher operator combinations. This appears to be
 straightforward for the exchange term (\ref{H1}) as long as one works in the
 rotated system $\Sigma^\prime$. We will perform the usual RPA
 decoupling\cite{BoT59} here.
\item The situation is more complex for the anisotropy term
  (\ref{anis}). Here we will develop a new decoupling scheme following
  the ideas of the Callen decoupling\cite{Cal63}. However the original
  Callen decoupling is not applicable, since the total spin is not a conserved quantity in our model. Therefore
  the decoupling has to be generalized.
\item There is a special rotation angle $\hat{\theta}$ in our model: If
  the coordinate system is rotated by this angle
  $\Sigma\rightarrow\hat{\Sigma}$ and the decoupling
  procedure is applied, the total
  magnetization $\sum_i S_{i{\hat{z}}}$ commutes with the Hamiltonian
  (\ref{hamiltonian}). It is easy
  to show that $\hat{\theta}$ is therefore the direction of the
  magnetization. Now the condition $[ \sum_i S_{i{\hat{z}}},H]_-=0$ gives an
  explicit expression for the magnetization angle.
\item Using the decouplings as well as the commutation property in the
  primed system we can solve the equation of motion and finally
  obtain the single magnon Green function as well as the norm of the
  magnetization $\langle S_{z^\prime}\rangle$. Therewith the problem
  will be solved.
\item One can further show, that the effect of the anisotropy
  can be interpreted instructively as an effective "anisotropy
  field". We will calculate the components of this field.
\end{enumerate}  
Lets now follow this program in more detail.
The rotation of the coordinate system is described by $\Sigma^\prime
=M\Sigma$, where $M$ is a rotation matrix. Due to the 
symmetry we may confine the rotation to the $zx$-plane without loss of
generality. This means that $y^\prime=y$ and that the polar angle
$\theta$ fully
characterizes the rotation.
\begin{eqnarray}
M=\left(\begin{array}{ccc}\cos\theta&0&-\sin\theta\\
0&1&0\\
\sin\theta&0&\cos\theta\end{array}\right).
\label{rota}
\end{eqnarray}
The $z^\prime$-axis of the new system $\Sigma^\prime$ is set to be parallel
to the magnetization direction. This gives:
\begin{eqnarray}
\langle
S_{x^\prime}\rangle=\langle S_{y^\prime}\rangle&=&0.
\label{ortho-zero2}
\end{eqnarray}
The magnetizations in the fixed system $\Sigma$ can now be read off from
Eq.~(\ref{rota}):
\begin{eqnarray}
\langle S_x\rangle = \sin\theta \langle S_{z^\prime} \rangle,\nonumber\\
\langle S_z\rangle = \cos\theta \langle S_{z^\prime} \rangle.
\label{projmag}
\end{eqnarray}
$\langle S_z\rangle$ is the magnetization component normal to the film
plane while $\langle S_x\rangle$ denotes the component parallel to
the film plane. $\langle S_{z^\prime}\rangle$, consequently, is the total
magnetization. Of course, the angle $\theta$ is a priori unknown.\\\\
Next we want to write down the equation of motion of the single magnon
Green function 
 $G_{ij}^\prime(E)=\langle\langle
{S}_i^{+\prime};{S}_j^{-\prime}\rangle\rangle$ defined in the new
system. Applying the transformation (\ref{rota}) to the Hamiltonian
(\ref{hamiltonian}) one readily finds:
\begin{eqnarray}
EG_{ij}^\prime(E)&=&\langle\left[S_i^{+\prime},S_j^{-\prime}\right]_-\rangle+\langle\langle\left[S_i^{+\prime},H\right]_;S_j^{-\prime}\rangle\rangle\nonumber\\
 &=&2\langle S_{z^\prime}\rangle\delta_{ij} -2J\sum_l^{\langle li\rangle}
 \left(\Gamma_{ilj}^\prime(E)-\Gamma_{lij}^\prime(E)\right)\nonumber\\
&&+K_2 (\cos^2\theta-\frac{1}{2}\sin^2\theta)\;\Gamma^{a\prime}_{ij}(E),\nonumber\\
&&+(B_{x0}\sin\theta+B_{z0}\cos\theta)G_{ij}^\prime(E)\nonumber\\
\nonumber\\
&&-(B_{x0}\cos\theta-B_{z0}\sin\theta)\langle\langle S_i^{z\prime};S_i^{-\prime}\rangle\rangle\nonumber\\ 
\nonumber\\
&&+2K_2\cos\theta\sin\theta\langle\langle S_i^{z\prime};S_i^{-\prime}\rangle\rangle\nonumber\\ 
\nonumber\\
&&+2K_2 \cos\theta\sin\theta \langle\langle {S_{i}^{z^\prime}}^2;S_i^{-\prime}\rangle\rangle\nonumber\\ 
\nonumber\\
&&-2K_2 \cos\theta\sin\theta \langle\langle S_{i}^{+\prime}S_{i}^{x^\prime};S_i^{-\prime}\rangle\rangle\nonumber
\end{eqnarray}
with
\begin{eqnarray}
 \Gamma_{mno}^\prime&=&\langle\langle S_m^{z\prime}
 S_n^{+\prime};S_o^{-\prime}\rangle\rangle\quad\mbox{and}\nonumber\\
\Gamma^{a\prime}_{ij}&=&\langle\langle
S_i^{+\prime}S_i^{z\prime}+S_i^{z\prime}S_i^{+\prime};S_j^{-\prime}\rangle\rangle. 
\label{eom}
\end{eqnarray}
The sum over $l$ runs over the nearest neighbors of site $i$.
The combination of trigonometric functions is a consequence of the
rotation (\ref{rota}). 
\\\\
To proceed the operator products in $\Gamma_{mno}^\prime$ and
$\Gamma^{a\prime}_{ij}$ have to be decoupled. For the nonlocal products 
in the former we choose a symmetric RPA decoupling $AB\rightarrow
\langle A\rangle B+ A\langle B\rangle $. Using Eq. (\ref{ortho-zero2})
which is valid in the primed system we find
\begin{equation}
S_m^{z\prime} S_n^{+\prime} \stackrel{RPA}{\longrightarrow} \langle
S_{z\prime} \rangle S_n^{+\prime}
\label{RPA}
\end{equation}
This is the same result as found in the original RPA
approach\cite{BoT59} for the "parallel limit". Let us
emphasize that this is only the case for the primed system
$\Sigma^\prime$. Thus the third point of our
program is done. \\\\
The crucial fourth step introduces a new approximation scheme for the
single ion anisotropy (\ref{anis}). It will be used to decouple the
higher Green function $\Gamma^{a\prime}_{ij}$ as well as to treat
operator combinations appearing in the commutator $[ \sum_i
S_{i{\hat{z}}},H]_-$. The latter is important for the fifth step of our calculation.
\\
1963 Callen introduced the Callen decoupling\cite{Cal63} which was
intended as an improvement to the RPA decoupling\cite{BoT59}. The
decoupling was performed at the Heisenberg exchange term
(\ref{H1}). Later Anderson and Callen\cite{AnC64} used this proposal to
treat the local anisotropy term (\ref{anis}) in the "parallel
limit". Both approaches are based on the fact that the total magnetization
commutes with the Hamiltonian, a condition that is not valid in our case.
Therefore the procedure has to be rederived and generalized. We will not
quite
adopt the procedure as presented in Ref. 9 but
rather use the main ideas.\\
The approximation makes use of the operator identity
\begin{equation}
S(S+1)-S_x^2-S_y^2-S_z^2=0
\label{identity}
\end{equation}
The spin operators work on the same site $i$, the site index is dropped
here for convenience.
Now the "zero" is added to the components of the spin operator,
$S_x, S_y, S_z$:
\begin{eqnarray}
S_{(x,y,z)}&=&S_{(x,y,z)}+\nonumber\\
&&\alpha_{(x,y,z)}\left(S(S+1)-S_x^2-S_y^2-S_z^2\right)
\label{add}
\end{eqnarray}
It is important to note that theses relations are identities for any
($\alpha_{(x,y,z)}$) only for
exact calculations. On the contrary the result of some standard
approximation procedure (e.g. of a symmetric mean field decoupling)
changes if one uses the right hand side of Eq. (\ref{add}) instead of
the left hand side. The results do depend now on the prefactors
($\alpha_{(x,y,z)}$). It was the idea of Callen to use this degree of
freedom to improve approximations. We will follow the proposal of Callen
here and adjust the parameters in a way that interpolates between zero
temperature and Curie temperature requirements. The explicit
calculation is given in Appendix A. It gives:
\begin{eqnarray}
\alpha_{(x,y,z)}&=&\frac{\langle S_{(x,y,z)}\rangle}{2S^2}
\label{prefact}
\end{eqnarray}
Now we want to decouple operator combinations like
$S_xS_z+S_zS_x,\,S_yS_z+S_zS_y$, or $S_xS_y+S_yS_x$ which appear in the
equation of motion (\ref{eom}) in the higher Green function
$\Gamma^{a\prime}_{ij}$.
Thereto we replace the single operators by the right hand side of
Eq. (\ref{add}) and perform a symmetric decoupling procedure at the
resulting expressions. Since the prefactors $\alpha_{(x,y,z)}$ are small
quantities we neglect terms of the order $\alpha^2$. For example the result for the
operator combination $S_yS_z+S_zS_y$ is thus given by:
\begin{eqnarray}
\label{A.C.-compl}
S_yS_z+S_zS_y&\stackrel{A.C.}{\longrightarrow}&2\langle S_y\rangle S_z+
2\langle S_z\rangle S_y\\
&&-2\frac{\langle S_y\rangle}{2S^2}\Big(\langle S_xS_z+S_zS_x\rangle
S_x\nonumber\\
&&\qquad+\langle
S_yS_z+S_zS_y\rangle S_y+2\langle S_z^2\rangle S_z\Big)\nonumber\\
&&-2\frac{\langle S_z\rangle}{2S^2}\Big(\langle S_xS_y+S_yS_x\rangle S_x\nonumber\\
&&\qquad+2\langle
S_y^2\rangle S_y+\langle S_zS_y+S_yS_z\rangle S_z\Big).\nonumber
\end{eqnarray}
For the other operator combinations analog expressions are found.\\
 By virtue of relation (\ref{ortho-zero2}) the
result can be simplified if the decoupling is performed in the rotated
coordinate system $\Sigma^\prime$. We show in Appendix B that the
following relations are fulfilled in $\Sigma^\prime$:
\begin{eqnarray}
\langle S_{a^\prime}S_{b^\prime}+S_{b^\prime}S_{a^\prime}\rangle&=&0,
\label{prime-rel}
\end{eqnarray}
where $a^\prime$ and $b^\prime$ are two different subscripts out of
$(x^\prime,y^\prime, z^\prime)$.
This, together with Eq. (\ref{ortho-zero2}), finally gives the decoupling
within the parallel system $\Sigma^\prime$:
\begin{eqnarray}
S_{x^\prime}S_{z^\prime}+S_{z^\prime}S_{x^\prime}&\stackrel{A.C.}{\longrightarrow}&
0\nonumber\\
S_{y^\prime}S_{z^\prime}+S_{z^\prime}S_{y^\prime}&\stackrel{A.C.}{\longrightarrow}&
2\langle S_{z^\prime}\rangle\left(1-\frac{\langle
    S_{y^\prime}^2\rangle}{S^2}\right)\cdot S_{y^\prime}\nonumber\\
S_{x^\prime}S_{z^\prime}+S_{z^\prime}S_{x^\prime}&\stackrel{A.C.}{\longrightarrow}&
2\langle S_{z^\prime}\rangle\left(1-\frac{\langle
    S_{x^\prime}^2\rangle}{S^2}\right)\cdot S_{x^\prime}
\label{A.C.-prime}
\end{eqnarray}
Now the operator combination that appears in the Green function
$\Gamma^{a\prime}_{ij}$ in the equation of motion (\ref{eom}) can be
decoupled. Using Eq. (\ref{A.C.-prime}) as well as the identity
(\ref{identity}) one finds: 
\begin{equation}
S_{+^\prime}S_{z^\prime}+S_{z^\prime}S_{+^\prime}\stackrel{A.C.}{\longrightarrow}
2\langle S_{z^\prime}\rangle C_{1}^\prime S_{+^\prime}\nonumber
\end{equation}
with
\begin{equation}
C_1^\prime=1-\frac{1}{2S^2}\left(S(S+1)-\langle
    S_{z^\prime}^2\rangle\right).
\label{S+Sz}
\end{equation}
\\\\ 
Using the decoupling procedures discussed up to now the higher Green
functions $\Gamma_{mno}^\prime$ and
$\Gamma^{a\prime}_{ij}$ in the equation of motion (\ref{eom}) can be treated. To treat the
other four terms we have to address the fifth point of our
program. Hence we will show in the following that for a certain angle
$\hat{\theta}$ the total magnetization $\sum_i S_i^{z^\prime}$ commutes with
the Hamiltonian (\ref{hamiltonian}). Furthermore an explicit expression for the magnetization
angle will be obtained.\\
Applying the rotation (\ref{rota}) to the Hamiltonian
(\ref{hamiltonian}) and using
the abbreviation $\gamma_1=\sin\theta$ and $\gamma_2=\cos\theta$
we find:
\begin{eqnarray}
\left[\sum_i S_i^{z^\prime},H\right]_-&=&\sum_i \left(\gamma_1 B_{z0}-\gamma_2
    B_{x0}\right)\cdot iS_{i}^{y^\prime}\nonumber\\
&&+K_2\gamma_1\gamma_2 \cdot
i\left(S_{i}^{y^\prime}S_i^{z^\prime}+S_i^{z^\prime}S_{i}^{y^\prime}\right)\nonumber
\end{eqnarray}
Now the last operator product is decoupled according to
Eq. (\ref{A.C.-prime}). This gives:
\begin{eqnarray}
\left[\sum_i S_i^{z^\prime},H\right]_-&=&\sum_i \bigg(\gamma_1 B_{z0}-\gamma_2
    B_{x0}\\
&&+2K_2\gamma_1\gamma_2 \langle S_{z^\prime}\rangle
  \Big(1-\frac{\langle S_{y^\prime}^2\rangle}{S^2}\Big)\bigg)\cdot iS_{i}^{y^\prime}\nonumber
\label{commutator}
\end{eqnarray}
Thus, in the framework of our approximation, the total magnetization
indeed commutes with the Hamiltonian if the term in brackets on the
right hand side is zero.
This has important consequences: All expectation values and Green
functions in
the rotated coordinate system $\hat{\Sigma}$ that do not conserve the spin are zero in the framework of our theory. This can be seen using, e.g., the
Lehmann representation of the Green function. In particular this applies
for $\langle S_{+\prime}\rangle$ and $\langle
S_{-\prime}\rangle$. 
 Therefore Eq. (\ref{ortho-zero2}) holds in this coordinate system which
 is thus found to be equivalent to the system $\Sigma^\prime$: 
 $\hat{\Sigma}=\Sigma^\prime$. Hence from Eq. (\ref{commutator}) follows simple condition for the magnetization angle $\theta$:
\begin{eqnarray}
0&\stackrel{!}{=}&\sin\theta B_{z0}-\cos\theta
    B_{x0}\nonumber\\
&&+2K_2\sin\theta\cos\theta \langle S_{z^\prime}\rangle C_1^\prime
 \label{angle1}
\end{eqnarray}
The expectation value $\langle S_y^2\rangle$ is already evaluated here
using the property of spin conservation in the primed
system. Eq. (\ref{angle1}) is our first important result.\\\\
Having calculated the magnetization angle now the norm of the
magnetization $\langle S_{z^\prime}\rangle$ has to be derived. This
turns out to be a straightforward task. 
Due to the property of spin
conservation in the primed system the Green functions in the last four
lines of the equation of motion (\ref{eom}) are identical to zero. The
remaining higher Green functions $\Gamma_{mno}^\prime$ and
$\Gamma^{a\prime}_{ij}$ can be decoupled by
Eqs. (\ref{RPA}) and (\ref{A.C.-prime}). Thus the equation of
  motion (\ref{eom}) can be solved after Fourier transformation. We finally obtain the
  single magnon Green function 
$G_\vec{q}^\prime(E)$,
\begin{eqnarray}
G_\vec{q}^\prime(E)&=&\frac{2\langle
  S_{z^\prime}\rangle}{E-E_\vec{q}^\prime}\nonumber\\
\mbox{with}&&\nonumber\\
E_\vec{q}^\prime&=&2\langle S_{z^\prime}\rangle
J(p-\gamma_\vec{q})+B\nonumber\\
B&=&B_{x0}\sin\theta+B_{z0}\cos\theta\nonumber\\
&&+K_2\left(\cos^2\theta-\frac{1}{2}\sin^2\theta\right)2\langle 
S_{z^\prime}\rangle C_1^\prime.
\label{GF-RPA}
\end{eqnarray}
The term $p$ denotes the coordination number, while $\gamma_\vec{q}$ is
a structural factor due to the Fourier transformation of the Heisenberg
exchange term. For the quadratic lattice chosen here it is given by:
\begin{eqnarray}
\gamma_\vec{q}=2\left(\cos aq_x+\cos aq_y \right),\nonumber
\end{eqnarray}
where $a$ is the lattice constant.
The trigonometric functions in Eq.~(\ref{GF-RPA}) are obviously a
consequence of the rotation. 
Knowing the Green function $G_\vec{q}^\prime(E)$ one can
calculate the desired expectation values in the primed system, i.e. the
total magnetization $\langle S_{z^\prime}\rangle$ and $\langle
S_{z^\prime}^2\rangle$ by a standard text book
procedure\cite{TaH62} finally ending up with a self consistent system of
equations. Before
discussing the results of our theory in more detail we want to offer an
instructive interpretation of the
work of the anisotropy in the framework of our approximation. This will
be the last point of our theory section.\\\\
The abbreviation $B$ in Eq.~(\ref{GF-RPA}) has the same effect on the
Green function as an
external field aligned parallel to the magnetization. Combining the
expression for $B$ with the magnetization angle, we may write down the
components of the effective field:
\begin{eqnarray}
B_x&=&B\sin\theta\nonumber\\
&=&B_{x0}-K_2\langle S_x\rangle \sin^2\theta C_1^\prime\nonumber\\
&=&B_{x0}+B_{xa}\nonumber\\
B_z&=&B\cos\theta\nonumber\\
&=&B_{z0}+2 K_2\langle S_z\rangle (1-\frac{1}{2}\sin^2\theta)C_1^\prime\nonumber\\
&=&B_{z0}+B_{za}.
\label{B-components}
\end{eqnarray}   
Obviously, the effective field may be written as a sum of the external
field and an "anisotropy field" $\vec{B}_a=(B_{xa},0,B_{za})$. The
anisotropy acts exactly like this field as far as the magnetization and
the magnon energies $E_\vec{q}$ are concerned.\\
In the next section we will
present the results of our theory and compare them with QMC data and
other approximations.

\section{Results and Discussion}
\begin{figure}
\epsfig{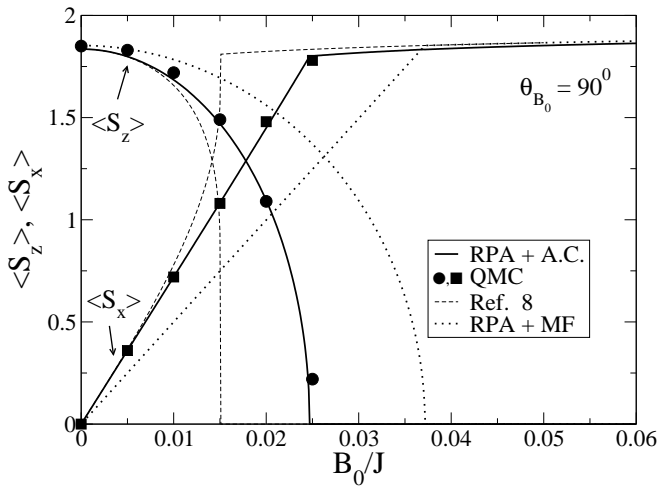}
\caption{The $x$- and the $z$-component of the magnetization $\langle
  S_x\rangle$, $\langle S_z\rangle$ as a function of the external field
  calculated with our RPA+A.C. approach (solid line), the approximation
  proposed in Ref. 8 (dashed line) and with a mean field
  decoupling of the anisotropy term (dotted line) in comparison with the
  QMC results from Ref. 6 (symbols). Parameters: $K_2=+0.02
  J$ and $k_BT=2J\approx 0.32k_BT_c, S=2.$} 
\label{an4}
\end{figure} 
We start our discussion with a comparison of our results to QMC data of
Ref. 6, which are free of systematic errors. Fig.~\ref{an4} shows the
results for the field induced reorientation transition at finite
temperatures. A positive anisotropy constant
($K_2>0$) and an external field parallel to the film plane
($\theta_{B_0}=90^\circ$) are considered. The external field is
applied perpendicular to the easy direction of the magnetization, a
situation representing a severe test for our theory.
 The components of the magnetization $\langle
S_x\rangle$ and $\langle S_z\rangle$ are shown as functions of the
external field. 
For zero field the magnetization is aligned parallel to
the easy axis. It is fully rotated into the film plane at the reorientation
field $B_{r0}$. We display the results of our calculations (RPA+A.C. -
solid lines) and the QMC results of
Ref.~6 (symbols). Additionally the results of two other theories are shown for
comparison: For the dotted line the anisotropy term (\ref{anis}) is
treated by simple mean field decoupling:
\begin{equation}
{S_i^z}S_i^z\stackrel{MF}{\longrightarrow}2\langle S_z\rangle S_i^z\nonumber
\end{equation}
The dashed line shows the proposal of Ref. 8. Here the
operator combination $S_i^+S_i^z+S_i^zS_i^+$ is decoupled as in
the parallel limit treated in the original paper of Anderson and Callen
\cite{AnC64}:
\begin{equation}
S_i^+S_i^z+S_i^zS_i^+\rightarrow 2\langle
S_z\rangle\left(1-\frac{1}{2S^2}(S(S+1)-\langle S_z^2\rangle)\right).\nonumber
\end{equation}
The exchange term in all model calculations is treated by an RPA decoupling.\\
\begin{figure}[t*]
\epsfig{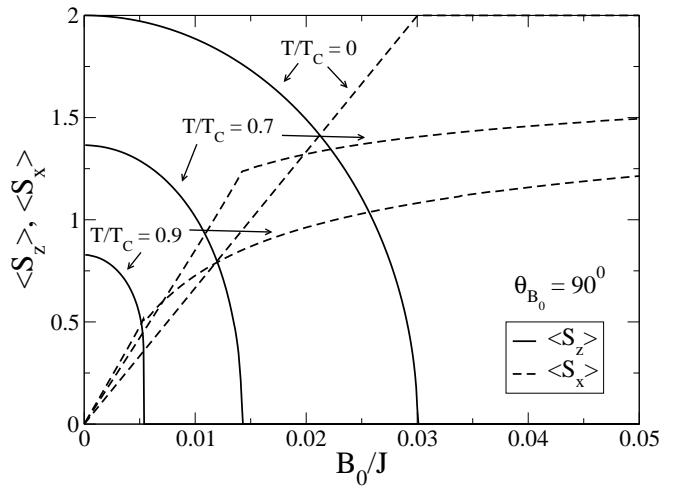}
\caption{The z-component $\langle S_z\rangle$ (solid lines) and the
  x-component $\langle S_x\rangle$ (dashed line) as a function of the
  external field $B_{x0}$ applied within the film plane.  Further
  parameters: $K_2=0.01 J,\,S=2$, $k_BT_c$ equals $5.75 J$.\cite{foot3}}  
\label{an5}
\end{figure}
The results of our theory (solid lines) are in excellent agreement with the QMC
data. We achieved even quantitative agreement for all magnetization
angles $\theta$. The quality of the approximation indeed turns out to be
the same for all angles $\theta$, which was the aim of this paper. 
Our approach is clearly superior to the
approximation proposed in Ref.~8, to the
mean field decoupling, and to all other approximations shown in
Fig. 11 of Ref.~6. Fig. \ref{an4} visualizes our main result, namely that we
succeeded to develop a theory for the extended Heisenberg model (\ref{hamiltonian})
that is as accurate as the RPA for the model
(\ref{H1}). In the following we will discuss some additional features of
the reorientation transition.
\\
In Fig.~\ref{an5} the temperature dependence of the transition is analyzed. Again, the components of the
magnetization are displayed as a function of the external field, which
is applied in the film plane, perpendicular to the easy direction.
The calculations were performed for
three different temperatures. Since the system is not saturated at
finite temperatures, the
total magnetization increases with the external field. This is seen best
after the reorientation
($B_{0}>B_{0r}$), where only one component of the magnetization
($\langle S_x\rangle$) is present.
\begin{figure}[t*]
\epsfig{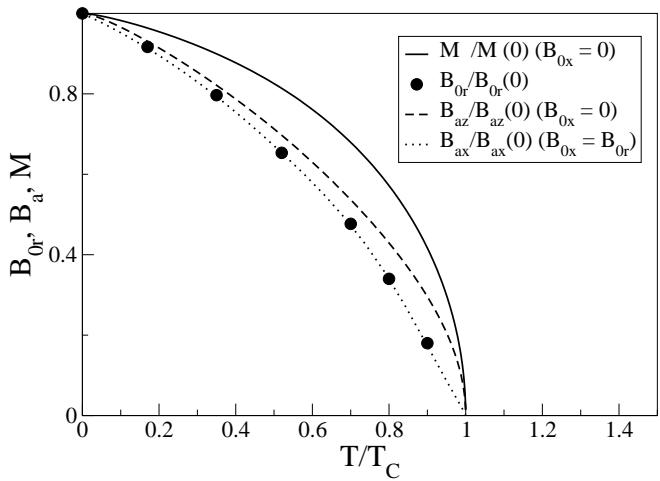}
\caption{The magnetization, the reorientation field $B_{0r}$ (dots) and the anisotropy field
  $B_a$ for $\theta=0^\circ$ (dashed line) and $\theta=90^\circ$
(dotted line) as a function of temperature. All quantities are scaled to
their zero temperature value. The other parameters are as in
Fig.~\ref{an5}. For Comparison scaled magnetization curves are added
(solid line).}  
\label{an6}
\end{figure}
For higher temperatures the transition as a function of the external
field becomes sharper. The reorientation field $B_{0r}$ decreases faster
with temperature than the zero field magnetization, reflecting the fact
that the anisotropy becomes less important at
higher temperatures. Another
interesting feature is that the $x-$component increases linearly with
the external field until the reorientation field is reached. This holds
for all temperatures and is
qualitatively different from the approximation proposed in
Ref.~8. Qualitatively, this feature is also found in mean-field theory
as can be seen in Fig. 1 of Ref.~14.\\
\begin{figure}[t]
\epsfig{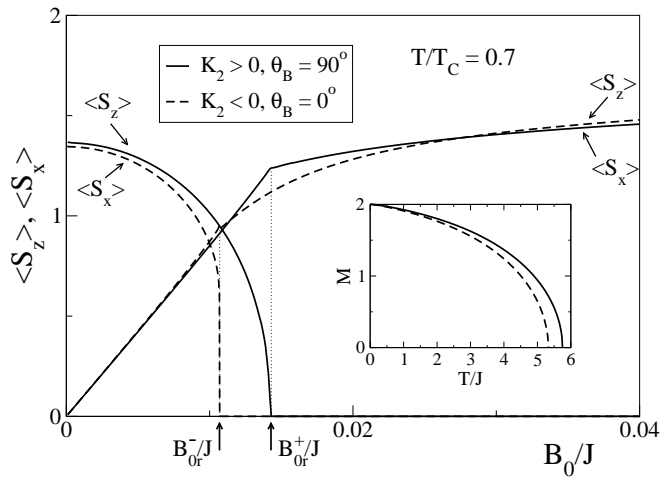}
\caption{The $z-$ and $x-$component of the magnetization as a function of
the external field. Results for positive $K_2=0.01 J$ (solid lines) and
negative $K_2=-0.01J$ (dashed lines) are shown, $T=0.7T_c$. The small arrows
highlight the position of the reorientation fields $B_{0r}^+$ for positive
 and $B_{0r}^-$ for negative  anisotropy. The inset shows the
magnetization curves for $B_0=0$. The other parameters are as in Fig.~\ref{an5}.}
\label{an7}
\end{figure}
The observed behavior 
follows directly from Eq. (\ref{angle1}). Since the external field is
applied parallel to the film plane  one finds for the
x-component of the magnetization $\langle S_x\rangle$:
\begin{eqnarray}
\langle S_x\rangle&=&\sin\theta\,\langle S_{z^\prime}\rangle\nonumber\\
&=& \frac{B_{x0}}{2 K_2 C_1^\prime(T)}.
\label{sinus}
\end{eqnarray} 
Since $C_1^\prime$ (Eq.\ref{GF-RPA}) increases with temperature,
 the slope of $\langle S_x\rangle$ is steeper
for higher temperatures.
Additionally, Eq.~(\ref{sinus}) determines the reorientation
field $B_{0r}$. We find: 
\begin{eqnarray}
B_{0r}(T)&=&2K_2\langle S_{z^\prime}\rangle(T)\; C_1^\prime(T).
\label{reorient2}
\end{eqnarray}
The fast decay of the reorientation field with temperature as compared
to the magnetization is also due to the temperature dependence of $C_1^\prime$.\\ 
This can also be seen in Fig.~\ref{an6}, where we considered the temperature
dependence of the system in detail. We plotted the norm of the components of the
anisotropy field (\ref{B-components}) as well the reorientation field (circles)
and the magnetization (solid line) as a function of temperature. All quantities are
scaled to their zero temperature value. The anisotropy fields are plotted at
their maxima, i.e. at $\theta=0^\circ$ for $B_{az}$ (dashed line) and at
$\theta=90^\circ$ for $B_{ax}$ (dotted line). The temperature dependence of the
anisotropy fields (\ref{B-components}) as well as of the reorientation
field (\ref{reorient2}) are determined by the factor $\langle
S_{z^\prime}\rangle(T)\;C_1^\prime(T)$. Thus this quantities have nearly
the same temperature dependence and their slopes are steeper than that of
the magnetization $\langle
S_{z^\prime}\rangle(T) $ alone.   
\\ 
Very similar results are found for the easy plane case ($K_2<0,\,\vec{B}_0\|z$).
In Fig.~\ref{an7} we compare both cases of a reorientation transition.
Solid lines show the transition for an easy axis system, dashed lines
denote the easy plane case.  
In the inset, the respective magnetization curves $M(T)=\langle
S_{z^\prime}\rangle(T)$ are plotted for zero external field. The Curie
temperature and the magnetization at finite temperatures are somewhat
smaller for the easy plane system. A reduced magnetization leads to
a reduced reorientation field (see Eq. (\ref{reorient2})). This explains the differences
between both cases concerning the reorientation transition as seen in
the main panel.  
\section{Conclusions and Outlook}
In this paper we addressed the magnetic reorientation transition in a
Heisenberg monolayer as a
function of the external field and temperature.
The basis of our approach is a transformation of the
Hamiltonian into a coordinate system $\Sigma^\prime$ (with the
$z^\prime$ axis parallel to the magnetization) as well as a
generalized Anderson-Callen decoupling procedure. Compared to the bare
Heisenberg Hamiltonian (\ref{H1}), the problem is more
complicated, since the total spin is not conserved. However, this complication turns out to be less serious, as it
can be shown that the total spin is a conserved quantity in the
framework of our approximation, if a appropriate quantization axis is
chosen. This fact can be used to calculate the
magnetization angle as well as to solve the equation of motion for the
single magnon Green function. It was further shown that the effect of the anisotropy can be
described by an effective "anisotropy field".\\  
Our results show a strikingly quantitative agreement with the QMC
data of Ref.~6 yielding a significant improvement over all other
decoupling schemes discussed so far (see e.g. Ref. 6).  
The main practical virtue of the new approach is that calculations can be
performed as accurate as with QMC but much
faster.\\ The theory can be generalized to a multilayer system and can thus
be used for cases where QMC calculations are not feasible any
more (e.g. thicker films). It should therefore be used to analyze the magnetic reorientation transition
as a function of the film thickness as found in many transition metal
films (see e.g. Ref. 4).\\
Due to its accuracy and convenience the theory shall further be used for
a quantitative analysis of
ferromagnetic resonance (FMR) experiments\cite{FMA97,FLB00}. The decisive feature for the
interpretation of a FMR experiment is the dependence of the $q=0$ spin
wave mode $E_{\vec{q}=0}$ on the external field $\vec{B}_0$. The
function $E_{\vec{q}=0}(\vec{B}_0)$ can be easily calculated in our theory
for any direction of the external
field. This opens the possibility to extract the microscopic anisotropy
constant $K_2$ directly from FMR experiments.\\

\section*{Acknowledgments}
This work is supported by the Deutsche Forschungsgemeinschaft within
the Sonderforschungsbereich 290.

\section*{Appendix A}
In this Appendix we derive the prefactors of Eq. (\ref{add}). 
We follow the philosophy of Callens paper\cite{Cal63} and
calculate the prefactors as an interpolation between low and high temperatures. Lets start
with the former limit $T\approx 0$:\\
Starting point is Eq. (\ref{add}). We will consider expectation
values instead of operators and transform the resulting expression as:
\begin{eqnarray}
\langle S_z\rangle &=& \langle S_z\rangle+\alpha_z\langle
S(S+1)-S_x^2-S_y^2-S_z^2\rangle\nonumber\\
&=& \langle S_z\rangle+\alpha_z\langle
S(S+1)-S_{x^\prime}^2-S_{y^\prime}^2-S_{z^\prime}^2\rangle\nonumber\\
&=& \langle S_z\rangle\nonumber\\
&&+\alpha_z\langle
S(S+1)-S_{z^\prime}-S_{-\prime}S_{+\prime}-S_{z^\prime}^2\rangle
\label{start}
\end{eqnarray}
The primed terms are quantities of the rotated system which is aligned
parallel to the magnetization.
Now the expectation values of the right hand side are approximated by
their zero temperature values
\begin{eqnarray}
\langle S_{z^\prime}\rangle &\stackrel{T\rightarrow 0}{\longrightarrow}&S\nonumber\\   
\langle S_{z^\prime}^2\rangle &\stackrel{T\rightarrow
  0}{\longrightarrow}&S^2
\end{eqnarray}
This gives
\begin{eqnarray}
\langle S_z\rangle\approx S\cos\theta +\alpha_z\langle
-S_{-\prime}S_{+\prime}\rangle
\end{eqnarray}
If $\alpha_z$ is set to zero, the left hand side and the right hand side
of Eq. (\ref{start}) are approximated on the same level, i.e. the
expectation values at low temperatures are replaced by their zero
temperature value. However
 one can even improve the approximation for the left
hand side by choosing $\alpha_z$ adequately. The choice 
\begin{equation}
\alpha_z(T\approx0)\stackrel{!}{=}\frac{\cos\theta}{2S}
\label{T0}
\end{equation} 
recovers the free spin wave result
\begin{eqnarray}
\langle S_z\rangle\approx \cos\theta\left(S-\frac{1}{2S}
\langle S_{-\prime}S_{+\prime}\rangle \right)
\end{eqnarray}
On the other hand for high temperatures $T>T_c$ the left hand side of
Eq. (\ref{start}) has to vanish. This can be assured by the choice
\begin{equation}
\alpha_z(T>T_c)\stackrel{!}{\sim}\cos\theta \langle S_{z^\prime}\rangle.
\label{TTc}
\end{equation}
Combining the settings (\ref{T0}) and (\ref{TTc}) one ends up with
Eq. (\ref{prefact}):
\begin{equation}
\alpha_z\stackrel{!}{=}\frac{\cos\theta \langle
  S_{z^\prime}\rangle}{2S^2}=\frac{\langle S_z\rangle}{2S^2}.\nonumber
\end{equation}
Analog calculations lead to the prefactors $\alpha_x$ and $\alpha_y$. 

\section*{Appendix B}
Here we want to derive the relation (\ref{prime-rel}). Starting point is
Eq. (\ref{ortho-zero2}). First we want to calculate the expectation
value
$\langle S_{x^\prime}S_{y^\prime}+S_{y^\prime}S_{x^\prime}\rangle$.
Using the decoupling (\ref{A.C.-compl}) together with
Eq. (\ref{ortho-zero2}) one finds
\begin{eqnarray}
\langle
S_{x^\prime}S_{y^\prime}+S_{y^\prime}S_{x^\prime}\rangle&\rightarrow&
\langle S_{x^\prime}\rangle {\cal A}+\langle S_{y^\prime}\rangle {\cal B}\nonumber\\
&=&0
\label{xy}
\end{eqnarray}
The terms ${\cal A}$ and ${\cal B}$ are given by the decoupling procedure
(\ref{A.C.-compl}).
This is one of three equations that have to be derived to prove relation
(\ref{prime-rel}).
Next we want to treat $\langle
S_{x^\prime}S_{z^\prime}+S_{z^\prime}S_{x^\prime}\rangle$.
Using the decoupling rule (\ref{A.C.-compl}) as well as the result
(\ref{xy}) one finds:
\begin{eqnarray}
S_{x^\prime}S_{z^\prime}+S_{z^\prime}S_{x^\prime}&\rightarrow&
2\langle S_{z^\prime}\rangle S_{x^\prime}\nonumber\\
&&-2\frac{\langle S_{z^\prime}\rangle}{2S^2}\Big(2 \langle
  S_{x^\prime}^2\rangle S_{x^\prime}\nonumber\\
&&+\frac{1}{2}\langle
S_{x^\prime}S_{z^\prime}+S_{z^\prime}S_{x^\prime}\rangle
S_{z^\prime}\Big)\nonumber
\end{eqnarray}
Thus it follows for the expectation value
\begin{eqnarray}
\langle S_{x^\prime}S_{z^\prime}+S_{z^\prime}S_{x^\prime}\rangle&\rightarrow&
2\langle S_{z^\prime}\rangle \langle S_{x^\prime}\rangle\nonumber\\
&&-2\frac{\langle S_{z^\prime}\rangle}{2S^2}\Big(2 \langle
  S_{x^\prime}^2\rangle \langle S_{x^\prime}\rangle\nonumber\\
&&+\frac{1}{2}\langle
S_{x^\prime}S_{z^\prime}+S_{z^\prime}S_{x^\prime}\rangle
\langle S_{z^\prime}\rangle\Big)\nonumber\\
&=&-\frac{\langle S_{z^\prime}\rangle}{2S^2}\left(\langle
S_{x^\prime}S_{z^\prime}+S_{z^\prime}S_{x^\prime}\rangle
\langle S_{z^\prime}\rangle\right)\nonumber\\
\mbox{therefore:}\nonumber\\
0&=&\langle
S_{x^\prime}S_{z^\prime}+S_{z^\prime}S_{x^\prime}\rangle \left(1+\frac{\langle
    S_{z^\prime}\rangle^2}{2S^2}\right)\nonumber\\
\mbox{and}\nonumber\\
0&=&\langle
S_{x^\prime}S_{z^\prime}+S_{z^\prime}S_{x^\prime}\rangle
\label{zx}
\end{eqnarray}
The equation 
\begin{equation}
\langle S_{y^\prime}S_{z^\prime}+S_{z^\prime}S_{y^\prime}\rangle=0
\label{yz}
\end{equation}
is derived in a analog way.
Eqs. (\ref{xy}), (\ref{zx}), and (\ref{yz}) prove relation (\ref{prime-rel}).


\begin{thebibliography}{10}
\bibitem{foot3}
These are the same parameters as in Ref. 6. Note that the definition of
$J$ in Ref. 6 differs from ours by a factor $1/2$ after a misprint in
Eq.~(11) of Ref. 6 is corrected.
\end{thebibliography}
\end{document}